\documentclass[twocolumn,aps,prl,groupedaddress]{revtex4-1}
\usepackage{graphicx}
\usepackage{color}
\usepackage{amsmath}
\usepackage{enumitem}
\usepackage{amssymb}
\usepackage{hyperref}

\newcommand{\be}{\begin{equation}}
\newcommand{\ee}{\end{equation}}

\newcommand{\bea}{\begin{eqnarray}}
\newcommand{\eea}{\end{eqnarray}}

\newcommand{\p}{\partial}

\newcommand{\la}{\left\langle}
\newcommand{\ra}{\right\rangle}

\newcommand{\lp}{\left(}
\newcommand{\rp}{\right)}
\newcommand{\sgn}{{\rm sgn\,}}

\renewcommand{\vec}[1]{{\bf #1}}

\newcommand{\addQ}[1]{\textcolor{red}{#1}}

\begin{document}
\title{
Particle Collisions and Negative Nonlocal Response of Ballistic Electrons}


\author{Andrey Shytov$^{a}$, Jian Feng Kong$^{b}$, Gregory Falkovich$^{c}$, Leonid Levitov$^{b}$}
\address{$^{a}$School of Physics, University of Exeter, Stocker Road, Exeter EX4 4QL, United Kingdom}
\address{$^{b}$Massachusetts Institute of Technology, Cambridge, Massachusetts 02139, USA}
\address{$^{c}$Weizmann Institute of Science, Rehovot 76100, Israel}
\date{\today}
\begin{abstract}
An electric field that builds in the direction against current, 
known as negative nonlocal resistance, arises naturally in viscous flows and is thus often taken as a telltale of this regime. Here we predict negative resistance for the ballistic regime, wherein the ee collision mean free path is greater than the length scale at which the system is being probed. Therefore, negative resistance alone does not provide strong evidence for the occurrence of the hydrodynamic regime; it must thus be demoted from the rank of a smoking gun to that of a mere forerunner. Furthermore, we find that negative response is log-enhanced in the ballistic regime by the physics related to the seminal Dorfman-Cohen log divergence due to  memory effects in the kinetics of dilute gases. 
The ballistic regime therefore offers a unique setting for exploring these interesting effects due to electron interactions.
\end{abstract}
%


\maketitle

Electron interactions 
can alter transport characteristics of solids in a variety of interesting ways\cite{LifshitzPitaevsky_Kinetics}. 
In particular, 
electron systems in which momentum-conserving ee collisions dominate transport 
are expected to exhibit collective hydrodynamic flows\cite{gurzhi63,dejong_molenkamp,jaggi91,damle97}. 
Viscous electron fluids can harbor 
interesting collective behaviors 
akin to those of classical fluids\cite{muller2009,andreev2011,forcella2014,tomadin2014,sheehy2007,fritz2008,narozhny2015,cortijo2015,LF,HG2}.  
Manifestations of electron hydrodynamics, predicted theoretically, provide guidance to experiments that attempt to demonstrate this regime\cite{bandurin2015,crossno2016,moll2016}. One such manifestation, discussed recently\cite{LF,bandurin2015}, is the ``negative resistance'' response i.e. current-induced electric field that builds in the direction against the applied current. In Ref.\cite{LF} negative resistance was predicted to arise naturally as the rate of momentum-conserving collisions exceeds the rate of momentum-relaxing collisions and the system transitions from the ohmic regime to the hydrodynamic regime. 
In Ref.\cite{bandurin2015} its observation was used as a signature of the hydrodynamic regime, 
taking it for granted that negative resistance is a fingerprint of the hydrodynamic regime.  
However, so far the smoking-gun status of this response has not been critically analyzed. 
 
\begin{figure}[t]
\begin{center}
\includegraphics[width=0.41\columnwidth]{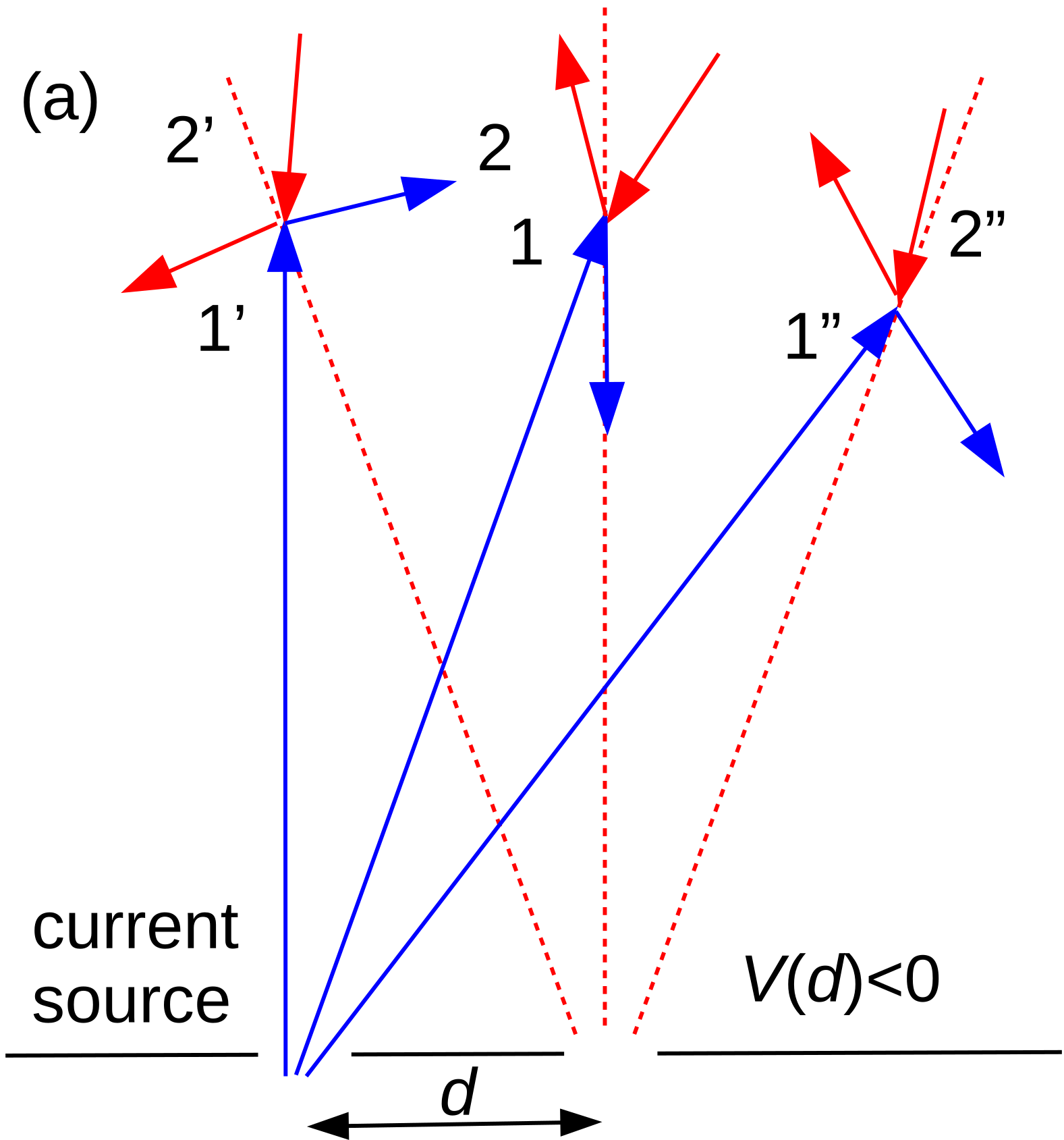} 
\ 
\includegraphics[width=0.56\columnwidth]{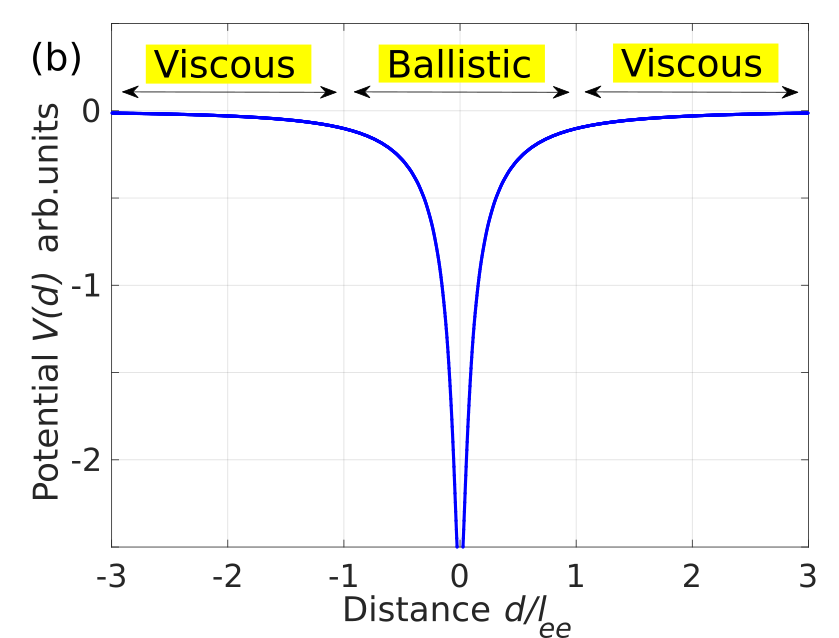}
 \end{center}
\caption{Particles injected into an electron system from a current source 
(blue) undergo collisions with particles in the system bulk (red). The change in particle distribution is detected by  a voltage probe at a distance 
$d$ from the source, which measures particle flux into a contact at the boundary. The signal, dominated by ee interactions, is strongest at the distances smaller than the ee collision mean free path, $d\ll l_{\rm ee}$. Panel (a) illustrates the mechanism of negative response: 
collisions between injected particles $1,1',1''$ and background particles $2,2',2''$ prevent some of the latter ($2',2''$) from entering the probe. Panel 
(b) shows the predicted dependence of the probe potential vs. distance. } 
 \label{fig1}
 \vspace{-5mm}
\end{figure}

Here we show that negative resistance can occur not only in the hydrodynamic regime, when the ee collision mean free path $l_{\rm ee}$ is the smallest lengthscale in the system, but also in the ballistic regime, when $l_{\rm ee}$ is much greater than the lengthscales at which the system is being probed.  This behavior is illustrated in Fig.\ref{fig1}. As a result, negative resistance, taken alone, does not distinguish the hydrodynamic and ballistic regimes. 
%
Furthermore, the negative response value in the ballistic regime exceeds that in the hydrodynamic regime, which puts certain limitations on using this quantity as a diagnostic of hydrodynamics. However, the two regimes can be distinguished by 
the temperature and carrier density dependence of the response. As discussed below, the response strength grows with temperature in the ballistic regime and decreases in the viscous regime. Likewise, it shows different dependence on doping in the two regimes. 
These dependences, which are strikingly different in the two regimes, can provide guidance in delineating them in the existing\cite{bandurin2015,berdyugin2018,bandurin2018,braem2018} and future experiments. Negative resistance in the ballistic regime is supported by recent measurements in graphene and GaAs electron gases\cite{bandurin2018,braem2018}.

 \begin{figure}[t]
\begin{center}
\includegraphics[width=0.99\columnwidth]{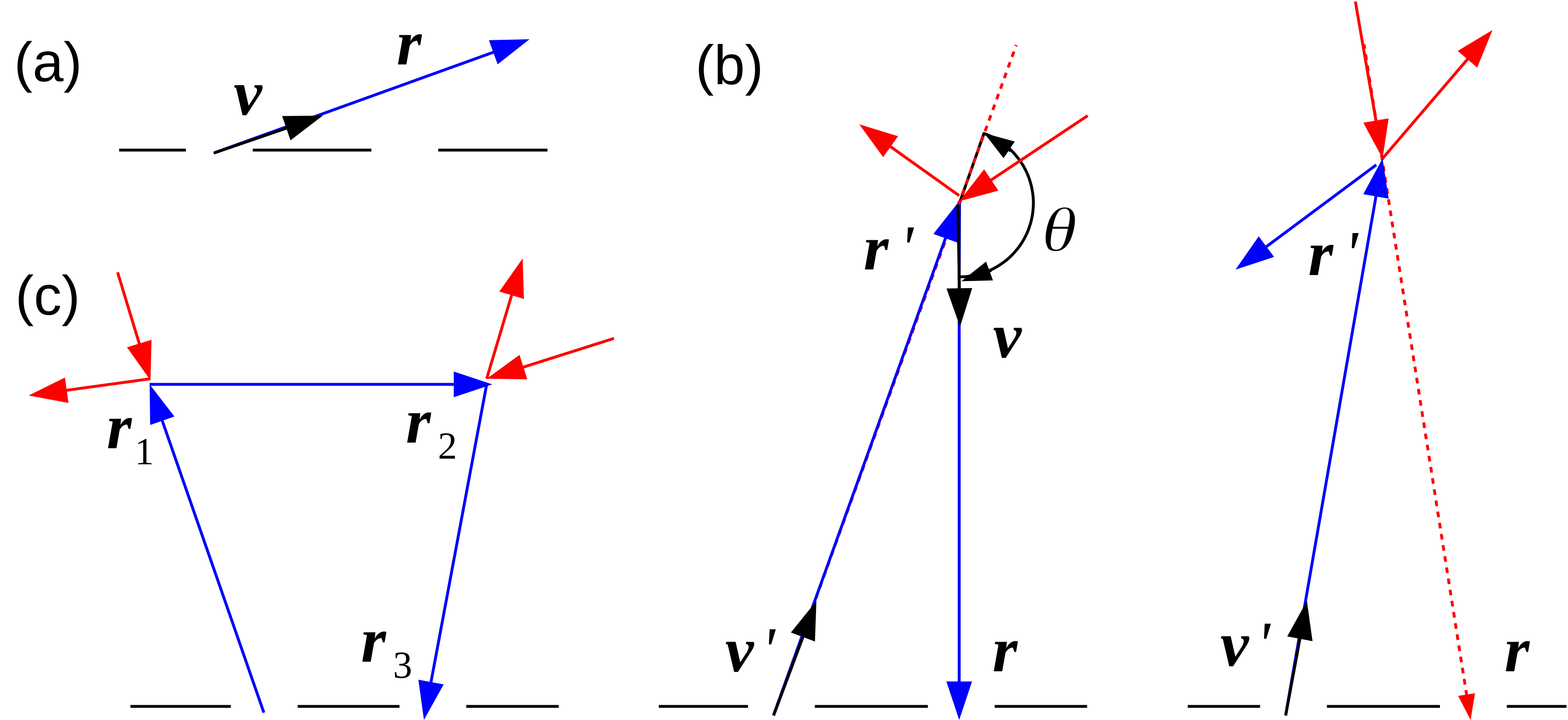} 
 \end{center}
\caption{Schematic illustration of different contributions to the nonlocal voltage response, arising in perturbation expansion of the solution of the transport equation, Eq.\eqref{eq:Boltzmann_perturation_expansion}, in the ee collisions rate. Panels (a), (b) and (c), illustrate the 1st, 2nd and 3rd terms, describing the result of $n=0$, $1$ and $2$ collisions, respectively. The dominant contribution, which is of a negative sign, arises from the 2nd 
contribution shown in (b), in which ambient carriers are scattered away and prevented from reaching the probe (see text).  
Such ``ghost'' processes
produce negative flux into the probe. 
These processes are pictured in Fig.\ref{fig1}a (particles $1',2'$ and $1'',2''$). 
}
 \label{fig2}
 \vspace{-5mm}
\end{figure}

The origin of negative resistance can be understood most easily by considering transport in the halfplane geometry wherein particles are injected from a point source placed at the boundary as shown in Fig.\ref{fig1}a. In this case there are two competing contributions to be considered. First, the injected particles, after colliding with the background particles, can be reflected into voltage probe which measures particle flux into the boundary. This produces a positive contribution to the measured voltage response. Second, the same collision processes also prevent some of the background particles from entering the probe, producing a negative contribution to the measured signal. Equivalently, this can be described as backscattering of a particle as a hole. We will see that the latter effect dominates, resulting in the net signal of a negative sign. 

Interestingly, when the ee mean free path $l_{\rm ee}$ is greater than the distance between the source and the probe $d$, all the lengthscales $d<r<l_{\rm ee}$ contribute equally to the response.
That is, the negative response is dominated by particles making a large excursion at $r>d$ before returning to the probe as a hole. In this case we find the behavior 
\be\label{eq:log_Vnegative}
V(d)\sim - J_0\gamma_{\rm ee}\ln\frac{l_{\rm ee}}{d}
,\quad d\ll l_{\rm ee}
\ee
 where $J_0$ is the injected current and $\gamma_{\rm ee}$ is the ee collision rate. As a function of distance, the response grows as $d$ decreases, diverging as $d/l_{\rm ee}\to 0$. This dependence is illustrated in Fig.\ref{fig1}b. In contrast, it falls off and becomes very small at large $d$, remaining negative in both the viscous 
regime $d\gg l_{\rm ee}$ and the ballistic regime $d\ll l_{\rm ee}$. As a function of distance to the probe, the negative response is 
stronger in the ballistic regime than  in the viscous regime. The log enhancement arises due to a large phase space of contributing trajectories, which make long excursions to the distances up to $l_{\rm ee}$ and then are scattered back to the probe as a hole, as illustrated in Figs.\ref{fig1} and \ref{fig2}b. 
 
The origin and behavior of the negative response bears 
a similarity to the seminal memory effects due to multiple correlated collisions in kinetic theory, discovered by  Dorfman and Cohen, and others\cite{dorfman1965,peierls}. This work made a surprising observation that virial expansion of the kinetic coefficients in gases breaks down due to multiple correlated collisions between two particles mediated by a third particle, which involve large excursions and log divergences similar to those found here. Manifestations of such memory effects, discussed so far, involved long-time power-law correlations in gases\cite{dorfman1970,dyakonov2005}. 
Here, instead of three correlated collisions, similar effects arise from a single collision, with the current source and voltage probe playing the role of two other collisions. One can therefore view the log enhancement in Eq.\ref{eq:log_Vnegative} as a direct manifestation of the memory effects predicted in kinetic theory. 
 
Our transport problem can be readily analyzed with the help of the quantum kinetic equation
\be\label{eq:B_eqn_with_source}
\lp 
\p_t+\vec v\nabla-I_{\rm ee}\rp \delta f(\vec r,\vec p)=J_{\vec r,\vec p}
,\quad
J_{\vec r,\vec p}=J_0\delta(\vec r)
.
\ee
Here $\delta f(\vec r,\vec p)$ describes particle distribution weakly perturbed near equilibrium. We assume $T\ll \epsilon_F$, in which case perturbed distribution is localized near the Fermi level and $\delta f(\vec r,\vec p)$ can be parameterized as a function on the Fermi surface through the standard ansatz 
\be\label{eq:harmonics}
\delta f(\vec r,\vec p)=-\frac{\p f_0}{\p\epsilon} \chi(\theta)
,\quad
\chi(\theta)=\sum_m \chi_m e^{im\theta}
,
\ee 
with $f_0$ the equilibrium Fermi-Dirac distribution and $\theta$ the angle parameterizing the Fermi surface. Due to cylindrical symmetry, the ee collision operator 
is in general diagonal in the angular harmonics basis (see below). The quantity $J_{\vec r,\vec p}$ represents a current source placed at $\vec r=0$. For conciseness, we ignore the angular anisotropy of the injected distribution. 

The general solution of this equation is given by a formal perturbation expansion in the collision term
\be\label{eq:Boltzmann_perturation_expansion}
\delta f(\vec r,\vec p)=DJ_{\vec r,\vec p}+DI_{\rm ee}DJ_{\vec r,\vec p}+DI_{\rm ee}DI_{\rm ee}DJ_{\vec r,\vec p}+...
\ee
where $D=(\delta+\vec v\nabla)^{-1}$ is the Liouville propagator. Here, to describe a steady state, an infinitesimal positive $\delta$ was added  in place of $\p_t$ to ensure that the steady-state response obeys causality. The collision processes described by this series are illustrated in Fig.\ref{fig2}. The first term represents particles
moving freely away from the source: 
\be\label{eq:free_motion}
\delta f_{1}(\vec r,\vec p)=\int_0^\infty dt \delta^{(2)}(\vec r-\vec v t)J_0
,
\ee 
where $t$ is an auxiliary time parameter arising from solving transport equations as $\delta f_{1}=\sum_{\vec k} e^{i\vec k\vec r}\frac{J}{\delta+i\vec k\vec v}=\sum_{\vec k}\int_0^\infty dt e^{i\vec k(\vec r-\vec v t)}J$. 
The particles described by Eq.\eqref{eq:free_motion} never make it to the probe (Fig.\ref{fig2}a). 
Other terms in Eq.\eqref{eq:Boltzmann_perturation_expansion} can also be evaluated in a similar manner. The second term describes injected particles scattered once by the background particles (Fig.\ref{fig2}b), giving
\be
\delta f_{2}(\vec r,\vec p)=\!\!\sum_{\vec r',t,t'}\!\! 
\delta^{(2)}(\vec r-\vec r'-\vec v t)\sigma(\theta)\delta^{(2)}(\vec r'-\vec v' t')J_0
,
\ee
where $\sum_{\vec r',t,t'}$ denotes $\int\limits_0^\infty \int\limits_0^\infty dt dt' \int d^2r'$, and 
the ``scattering crosssection'' $\sigma$ describes the change of the distribution due to a scattering event. The crosssection dependence vs. the angle between the incoming and outgoing velocities $\theta$ (see Fig.\ref{fig2}b) can be inferred from the form of the collision  operator $I_{\rm ee}$. For illustration, here we consider the simplest one-rate model of $I_{\rm ee}$ in which all nonconserved harmonics 
relax at equal rates,\cite{dejong_molenkamp,HG2} 
\be\label{eq:Iee_Molenkamp}
I_{\rm ee}\delta f=-\gamma_{\rm ee}\lp \delta f -2\hat{\vec p}\cdot\la \hat{\vec p}'\delta f'\ra_{\theta'}-\la\delta f'\ra_{\theta'}\rp
,
\ee
where 
the average $\la...\ra_{\theta'}$ is over $\vec p'$ angles; 
$\delta f$ and $\delta f'$ is a shorthand for $\delta f(\vec p,\vec r)$ and $\delta f(\vec p',\vec r)$, respectively. 
The last two terms in
Eq.\eqref{eq:Iee_Molenkamp}, which 
ensure momentum and particle number conservation,  give the angle dependence
\be\label{eq:sigma(theta)}
\sigma(\theta)=\gamma_{\rm ee}(1+2\cos \theta)
.
\ee
The two terms in this expression have very different meanings: the first, isotropic, term describes addition of an incident particle after its velocity is randomized by collision, the second term describes momentum recoil of the background particles as a result of scattering. 

Crucially, the crosssection $\theta$ 
dependence in Eq.\eqref{eq:sigma(theta)} is such that $\sigma$ is positive at small $\theta$ but is {\it negative} in an interval of size $2\pi/3$ which includes the scattering angle $\theta=\pi$. The contribution of this process to the flux into the probe is dominated by the values $\theta\approx \pi-O(d/r)$. This contribution originates from scattering processes at relatively large distances from the injector $r\gg d$, giving a negative value which is 
log-enhanced: 
\be
\delta V\sim J_0  \int_{d}^\infty \frac{d^2r'}{{r'}^2} 
e^{-r'/l_{\rm ee}}\sigma(\theta\approx\pi )
\sim - J_0\gamma_{\rm ee}\ln\frac{l_{\rm ee}}{d} 
.
\ee
The log factor is large in the ballistic regime $l_{\rm ee}\gg d$. 

The textbook estimate $\gamma_{\rm ee}\sim bR^* T^2/\epsilon_F^2$, where $R^*$ is the
effective Rydberg constant near $\epsilon_F$ and $b$ is a numerical factor of order unity,  indicates that the response grows with temperature ($T$) and decreases with carrier density ($n$). This is in contrast to the negative response in the hydrodynamic regime, which is proportional to viscosity and thus scales inversely with $\gamma_{\rm ee}$\cite{LF}. The opposite signs of the dependence vs. $T$ and $n$ 
may help distinguish the ballistic and viscous negative response. 

Higher-order terms  in Eq.\eqref{eq:Boltzmann_perturation_expansion} describe multiple scattering.  E.g. the third term gives a contribution to particle flux into the probe of the form
(Fig.\ref{fig2}c):
\be
J_0\gamma_{\rm ee}^2 \int \int \frac{d^2r_1d^2 r_2 e^{-L/l_{\rm ee}}}{|\vec r_1||\vec r_2-\vec r_1||\vec r_3-\vec r_2|} \sim \gamma_{\rm ee} J_0
,
\ee
where $L=|\vec r_1|+|\vec r_2-\vec r_1|+|\vec r_3-\vec r_2|$. This contribution
is non-divergent in the limit $d\ll l_{\rm ee}$, and thus is subleading to the second term by a log factor. 

By a similar dimensional argument one can show that $n$th order terms gives  contributions
\be
J_0\gamma_{\rm ee}^n\int ...\int \frac{d^2r_1d^2r_2...d^2 r_n}{|\vec r_1| |\vec r_1-\vec r_2|...|\vec r_n-\vec r_{n-1}|}
\sim \gamma_{\rm ee}^n\frac{ l_{\rm ee}^{2n}}{l_{\rm ee}^{n+1}}
\sim\gamma_{\rm ee}
.
\ee
This behavior of higher-order terms, featuring identical scaling with $\gamma_{\rm ee}$, simply means that perturbation expansion is ill-defined and cannot be used to evaluate the response outside the ballistic regime. 
As noted above, the log divergence of the second term and the power-law divergence of higher-order terms are 
related to the seminal divergences found in the breakdown of the virial expansion in kinetic theory due to memory effects in multiple correlated collisions\cite{dorfman1965,peierls}. 

We now proceed to show that the nonlocal resistance also remains negative outside the ballistic regime, that is at large distances $r\gg l_{\rm ee}$. 
To describe this regime we need to incorporate boundary scattering into the model. Momentum relaxation at the boundary is usually described by diffuse boundary conditions, leading to a cumbersome mathematical boundary value problem. Instead, to simplify the analysis, here we extend particle dynamics from the  halfplane to the full plane, and model momentum relaxation on the line $y=0$ through adding an additional term to the collision operator as
\be
I_{\rm ee} \to I_{\rm ee}+I_{\rm bd}
,\quad
I_{\rm bd} \delta f=-\alpha\delta(y)P' \delta f
.
\ee
Here $P'$ is a projection on the harmonics $m=\pm 1$: $P'\delta f=2\hat{\vec p}\cdot\la \hat{\vec p}' \delta f(\vec p')\ra_{\vec p'}$. 
The limit $\alpha\to\infty$ 
is expected to mimic the no-slip boundary conditions. Carrier distribution induced by an injector is described by 
\be\label{eq:Boltzmann_J0}
(\vec v\nabla-I_{\rm ee}+\alpha(\vec r)P')\delta f(\vec r,\vec p)=J_0\delta(\vec r)
.
\ee
The solution of this transport problem can be obtained in Fourier representation $\delta f(\vec r,\vec p)=\sum_{\vec k} e^{i\vec k\vec r} \delta f_{\vec k}(\vec r)$:
\be\label{eq:G(f0+df)}
( i\vec k\vec v -I_{\rm ee}+\hat\alpha P' ) \delta f_{\vec k}(\vec r) 
=J_0 
, \quad 
\la \vec k|\hat\alpha|\vec k'\ra=\alpha \delta_{k_1-k'_1}
, 
\ee
where the delta function $\delta_{k_1-k'_1}$ reflects translational invariance of the line $y=0$ in the $x$ direction.  

Next, we transform to the angular harmonics basis \eqref{eq:harmonics}. We formally solve Eq.\eqref{eq:G(f0+df)} by a perturbation series in $\alpha$:
\be\label{eq:df_main}
\left. | \delta f\ra=\lp G-G\hat\alpha G+G\hat\alpha G\hat\alpha G-...
\rp \left.|0\ra J_0
,
\ee
where, $G=1/(i\vec k\vec v 
-I_{\rm ee})$ is the free-space Green's function, $\left.|0\ra $ denotes the $m=0$ angular harmonic. For conciseness, we absorb $P'$ into $\hat\alpha$ and suppress the $\p f_0/\p\epsilon$ factor. 
The first term represents a solution of Eq.\eqref{eq:G(f0+df)} for a point source in free space and no momentum relaxation, $\alpha=0$. Other terms describe 
scattering at the line $y=0$. 
Because of $P'$ projection, every encounter with the line 
generates a contribution of the form $e^{i\theta}+e^{-i\theta}=2\cos\theta$. We can therefore replace Eq.\eqref{eq:df_main} by an equivalent free-space problem with a line source 
\be\label{eq:df_w_line_source}
(ivk\cos(\theta-\theta_k)-I_{\rm ee})\left. | \delta f\ra=
J_0(1+w_{k_1} 2\cos\theta) 
.
\ee
Here $\theta$ is the velocity angle and $\theta_k$ is the vector $\vec k$ angle, $k_1+ik_2=k e^{i\theta_k}$. 
The first term $1$ on the right hand side represents the original point source at $\vec r=0$; the terms $w_{k_1} e^{\pm i\theta}$ represent a source distributed on the $y=0$ line (no $k_2$ dependence). The weights $w_{k_1}$ are evaluated in the Supplemental Material. 

In the basis \eqref{eq:harmonics}, the transport problem \eqref{eq:df_w_line_source} is represented as a system of coupled equations 
\be\label{eq:coupled_eqs}
\frac{ikv}{2}\lp e^{i\theta_k} \chi_{m+1}+e^{-i\theta_k} \chi_{m-1}\rp+\gamma_m\chi_m=J_m 
,
\ee
where $\gamma_m$ are the eigenvalues of the operator $I_{\rm ee}$, which is diagonal in the basis \eqref{eq:harmonics}, and 
$J_m$ take values $J_0$ and $w_{k_1}J_0$ for $m=0,\pm 1$ and zero otherwise. 
Here we used the identity $\cos(\theta-\theta_k)=\frac12(e^{i(\theta-\theta_k)}+e^{-i(\theta-\theta_k)})$, interpreting the factors $e^{\pm i\theta}$ as shift operators $m\to m\mp 1$. 

In our one-rate model the eigenvalues of $I_{\rm ee}$ are $\gamma_m=\gamma_{\rm ee}$ for $|m|>1$, and zero otherwise. We will now show that in this case the coupled equations, Eq.\eqref{eq:coupled_eqs}, have a solution with the $m$ dependence of an exponential form
\be
\chi_{m}=e^{ -im\theta_k}\times \left\{\begin{array}{lr} c_1 (-iz)^{m-1}, & m>0 \\ c_0, & m=0 \\ c_{-1}(-iz)^{-(m+1)}, & m<0 \end{array}\right.
\ee
with $|z|<1$. Plugging it into Eq.\eqref{eq:coupled_eqs} with any $m\ne 0,\pm 1$ gives an algebraic equation $\frac{vk}2\lp z-z^{-1}\rp+\gamma_{\rm ee}=0$. This equation is solved by
\be
z=e^{-\lambda},\quad
\sinh\lambda=\frac{\gamma_{\rm ee}}{kv}
.
\ee
The $m=\pm 1$ and $m=0$ equations are
\be
c_0-iz c_{\pm 1}=e^{\pm i\theta_k}w_{k_1} \frac{2J_0}{ikv} 
,\quad
c_1+c_{-1}=\frac{2J_0}{ikv}
.
\ee
These equations give values 
\be
c_0=J_0\frac{2w_{k_1}\cos\theta_k+iz}{ikv}
,\quad
c_{\pm 1}=J_0\frac{z\mp 2 w_{k_1}\sin\theta_k}{ikvz}
.
\ee
The full distribution can now be evaluated by carrying out the sum over $m$. This gives a closed-form expression
\be\label{eq:chi(theta)}
\delta f_{\vec k}(\theta)=c_0+\frac{c_1 e^{i(\theta-\theta_k)}}{1+ize^{i(\theta-\theta_k)}}+\frac{c_{-1} e^{-i(\theta-\theta_k)}}{1+ize^{-i(\theta-\theta_k)}}
\ee
where the three terms represent the contributions of the harmonics $m=0$, $m>0$ and $m<0$, respectively. 

We model the voltage probe as a small slit 
which measures the incoming particle flux $F$ (see Fig.\ref{fig1}): 
\be\label{eq:V(r)_general}
V(d)=\frac{e w}{G}F,\quad
F=\int_{-\pi}^0\frac{d\theta}{2\pi}Dv\sin\theta \chi(r,\theta) 
,\quad
\ee
where the integration limits $-\pi<\theta<0$ select particles which are incident on the boundary. 
Here $w$ is the slit width, $e$ is electron charge, $G=(4e^2/h)(2w/\lambda_F)$ is 
the slit conductance, and $D$ is the density of states. 
Particles incident at an angle $\theta$ 
contribute to the flux with the weight $v\sin\theta$.  The voltage $V(d)$ does not depend on the slit width $w$, as expected. 

We emphasize that the voltage probe measures the incoming current flux rather than the current-induced potential or charge density change. Indeed the injected current gives rise to a space charge buildup in the system bulk. This space charge, due to quasineutrality, shifts local chemical potential. However, in a steady state, a change in the local chemical potential does not lead to a net current into the boundary and therefore does not contribute to the voltage signal measured by the probe.

We evaluate voltage on the probe, Eq.\eqref{eq:V(r)_general}, using the 
carrier distribution 
\eqref{eq:chi(theta)}, Fourier transformed to real  space. 
The flux for the distribution \eqref{eq:chi(theta)} can be  analyzed by summing the contributions of different harmonics with the help of the identity
\be
\int\limits_{-\pi}^0\frac{d\theta}{2\pi}v\sin\theta e^{im\theta}
=\left\{\begin{array}{lr} \frac{v}{\pi(m^2-1)}, 
& m\ {\rm even} \\ \pm \frac{iv}4, &  m\ {\rm odd}, m=\pm 1
 \\ 0, & m\ {\rm odd}, m\ne \pm1
\end{array}\right.
.
\ee
The resulting response, illustrated in Fig.\ref{fig1}b, is negative in both the ballistic and viscous 
regimes. It is more negative in the ballistic regime, $d\ll l_{\rm ee}$, 
than in the viscous 
regime, $d\gg l_{\rm ee}$. Therefore, the sign of the response does not distinguish between the two regimes. However, since in the ballistic regime the response scales as $\gamma_{\rm ee}$, whereas in the viscous 
regime it scales as $\gamma_{\rm ee}^{-1}$, the $T$ and $n$ 
dependences will be of opposite signs in the two cases, providing a clear signature that may help distinguish 
the two regimes. 

For monolayer graphene the negative response of ballistic electrons, derived above, is proportional to $\lambda_F\gamma_{\rm ee}\sim T^2/n$, 
decreasing with $n$ and growing with $T$. 
Yet, for  a viscous flow the 
response is proportional to $\eta/n^2$, where $\eta$ is viscosity. 
The estimate $\eta=nmv_Fl_{\rm ee}/4$ then predicts a density-independent response. Interestingly, the 
response measured in Ref.\cite{bandurin2015} decreases with $n$ and grows with $T$ at not-too-high temperatures, 
resembling the behavior expected for ballistic electrons. The vicinity resistance geometry therefore provides an ideal setting in which the effects of ee interactions in the ballistic regime can be explored.

Part of this work was performed at the Aspen Center for Physics, which is supported by National Science Foundation Grant No. PHY-1607611. We acknowledge support by A*STAR NSS (PhD) Fellowship (J. F. K.); the Minerva Foundation, ISF Grant 882, the RSF Project No. 14-22-00259 (G. F.); the Center of Integrated Quantum Materials under NSF Grant No. DMR-1231319; the MIT Center for Excitonics, the Energy Frontier Research Center funded by the U.S. Department of Energy, Office of Science under Award No. DE-SC0001088, and Army Research Office Grant No. W911NF-18-1-0116 (L. L.).



	\setcounter{equation}{0}
	\setcounter{figure}{0}
	\renewcommand{\thesection}{}
	\renewcommand{\thesubsection}{\arabic{subsection}}
	\renewcommand{\theequation} {S\arabic{equation}}
	\renewcommand{\thefigure} {S\arabic{figure}}

\subsection{Supplemental Material}

We are interested in the response of the system arising at the lengthscales comparable or smaller than the ee collisions mean free path. In this case, it is convenient to employ the so-called {\it quasi-hydrodynamic variables}, i.e. the microscopic quantities projected on the hydrodynamic subspace of the angular harmonics of particle distribution that do not relax through momentum-conserving ee collisions. 

To construct such a framework in the linear response regime, when the system is weakly perturbed about equilibrium, it is sufficient to work with a system linearized 
about the equilibrium state. Linearized transport equation defines a linear operator acting in the space of carrier distributions. Below we evaluate the Greens function, i.e. the resolvent of the linearized  transport operator, by the projection approach. 

Namely, we assume $T\ll \epsilon_F$, in which case the perturbed distribution is localized near the Fermi level 
and $\delta f(\vec r,\vec p)$ can be parameterized as a function on the Fermi surface through the standard ansatz $\delta f(\vec p)=-\frac{\p f_0}{\p\epsilon} \chi(\theta)$ with $f_0$ the equilibrium Fermi-Dirac distribution. Owing to cylindrical symmetry, the linearized collision operator $I_{\rm ee}$  is in general diagonal in the angular harmonics basis $\chi(\theta)=\sum_m \chi_m e^{im\theta}$: 
\be
\left. I_{\rm ee}|\chi\ra=\sum_m -\gamma_m \left.|m\ra\la m|\chi\ra
,\quad
\left.|m\ra=e^{im\theta}
\ee
where $\gamma_m$ are the relaxation rates for individual harmonics. 
The operator $I_{\rm ee}$ is hermitian with respect to the inner product $\la \chi_1|\chi_2\ra=\oint\frac{d\theta}{2\pi}\chi_1^\ast(\theta)\chi_2(\theta)$. 
In this notation, the one-rate collision operator used in the main text, Eq.\eqref{eq:Iee_Molenkamp}, is written as
\be\label{eq:Iee=1-P}
I_{\rm ee}(f)=-\gamma(f-Pf)
,\quad
P=\left. |0\ra \la 0|+|1\ra \la 1|+|-1\ra \la -1|\right.
.
\ee
We start with analyzing the linearized transport problem in free space
\be
(\vec v\nabla-I_{\rm ee})\delta f(\vec r,\vec p)=J(\vec r,\vec p)
.
\ee
The Green's function for this problem can described 
in Fourier representation $\delta f(\vec r,\vec p)=\sum_{\vec k} e^{i\vec k\vec r}\delta f_{\vec k}(\vec p)$. We have 
\be\label{eq:Greens_fcn}
\begin{split}
&\delta f_{\vec k}(\vec p)=\oint\frac{d\theta'}{2\pi}G_{\vec k}(\theta,\theta')
J_{\vec r}(\vec p')
,\quad
\\
&G_{\vec k}(\theta,\theta')=\la \theta \Big| \frac1{\gamma(1-P)+i\vec v\vec k}\Big|\theta'\ra
.
\end{split}
\ee
The quantities $\gamma(1-P)$ and $i\vec v\vec k$, viewed as operators in the space of angular harmonics, do not commute. Therefore, evaluating the Greens function in Eq.\eqref{eq:Greens_fcn} is in general a nontrivial exercise. One can simplify the task by projecting 
on the hydrodynamic subspace, i.e. the $m=0,\pm1$ harmonics. This subspace represents the target space of the projection operator $P$, defined in Eq.\eqref{eq:Iee=1-P}: $\left. P|m\ra=\left. |m\ra$. In doing so we define the projected  Greens function
\be
D=PGP
,\quad
G=\frac1{\gamma(1-P)+i\vec v\vec k}
.
\ee
The quantity $D$ has a number of advantages over $G$. First, it is a matrix of a finite size ($3\times3$) whereas $G$ is an infinite-rank matrix. Second, it encodes in a simple way all the information about the hydrodynamic modes originating from the conserved harmonics $m=0,\pm1$ (particle density and momentum density). And lastly, this quantity can be evaluated in a closed form by a $T$-matrix approach described below.

To evaluate $D$ we proceed in two steps. First, we evaluate the $3\times 3$ matrix $g=\gamma PG_0P$ where $G_0=1/(i\vec k\vec v\cos\tilde\theta+\gamma)$ is an auxiliary Green's function describing transport in which {\it all harmonics,} including $m=0,\pm1$, relax at a rate $\gamma$. Direct calculation gives matrix elements (here $m,m'=0,\pm 1$, $\Delta m=m-m'$):
\be\label{eq:g3x3}
g_{mm'}=\la \frac{\gamma e^{i(m-m')\theta}}{\gamma+i\vec k\vec v\cos\tilde\theta}\ra_\theta
\!\!
=\tanh\beta \frac{ e^{i\theta_k\Delta m} }{\lp ie^{\beta}\rp^{|\Delta m|}}
.
\ee
Here we introduced notation $\sinh\beta=\frac{\gamma}{kv}$ and defined $\tilde\theta=\theta-\theta_k$, the angle between particle velocity $\vec v$ and momentum $\vec k$. 

The matrix $D$ can now be expressed through the matrix $g$ 
by expanding the full Green's function as $G=1/(G_0^{-1}-\gamma P)=G_0+G_0\gamma PG_0+...$, which gives 
\be\label{eq:G_Tmatrix}
G=G_0+G_0TG_0
,\quad
T=\frac{\gamma P}{1-\gamma PG_0P}
.
\ee
To arrive at Eq.\eqref{eq:G_Tmatrix} we re-summed the series, expressing the result in terms of a $3\times 3$ matrix $T$ in a manner analogous to the derivation of the Lippmann-Schwinger $T$-matrix for quantum scattering with a finite number of `active' channels.
We note that $\gamma PG_0P$ is nothing but the matrix $g$ in Eq.\eqref{eq:g3x3}. 
Plugging the T-matrix expression for $G$, Eq.\eqref{eq:G_Tmatrix}, into $D=PGP$ 
and carrying out a tedious but straightforward matrix inversion we obtain a closed-form expression
\be\label{eq:D}
D=\frac{\gamma^{-1}g}{1-g}
=\frac{\sinh\beta}{\gamma} \!\lp\begin{array}{ccc} e^{\beta} & -i z_k &-e^{\beta} z_k^2\\ -i\bar z_k & e^{-\beta} & -i z_k \\ -e^{\beta} \bar z_k^2 & -i \bar z_k & e^{\beta} \end{array}\rp
,
\ee
where $z_k=e^{i\theta_k}$.

As a next step, we apply the above result to the transport problem in which momentum relaxation takes place at system boundary, modeled in the main text as transport in free space with momentum relaxation on a line $y=0$, Eq.\ref{eq:Boltzmann_J0}. Particle distribution, induced in system bulk by a point source positioned at the boundary, is described in Fourier representation by Eq.\eqref{eq:G(f0+df)}. 
Formal solution of this equation $\left. | \delta f\ra=\lp G-G\hat\alpha G+G\hat\alpha G\hat\alpha G-...
\rp \left.|0\ra J_0$, obtained by perturbation expansion in $\alpha$, Eq.\eqref{eq:df_main}, can be written in terms of the projected Green's function $D$ as follows: 
\be\label{eq:df}
\left.  | \delta f\ra=\lp G-G\hat\alpha D+G\hat\alpha D\hat\alpha D-...
\rp \left. |0\ra J_0
\ee
where 
we used the identity $\hat\alpha=P\hat\alpha P$ which follows from $PP'=P'P=P'$.
This simple structure, with the full Greens functions $G$ replaced by the projected functions $D$, arises because the scattering processes at the boundary affect only the $m=\pm1$ harmonics.

To evaluate the series in Eq.\eqref{eq:df} we note that the vertex $\hat \alpha$ conserves the momentum $x$ component but does not conserve the $y$ component. We therefore must integrate over $k_2$ independently in each block  $\hat\alpha D\hat\alpha$, keeping the value $k_1$ fixed throughout. 
Reinstating $P'$ in $\hat\alpha$ and
noting that $P'DP'$ eliminates the middle row and column in $D$, corresponding to $m=0$, 
we obtain
\be\label{eq:aDa}
\hat\alpha D\hat\alpha=\alpha^2\int\limits_{-\infty}^\infty \frac{dk_2}{2\pi} 
\frac{\sinh\beta}{\gamma}  \lp\begin{array}{cc} e^{\beta} &-e^{\beta} z_k^2\\ -e^{\beta}\bar z_k^2 & e^{\beta} \end{array}\rp
.
\ee
Integration over $k_2$, while somewhat cumbersome, can be performed in a closed form. It is convenient to nondimensionalize the 
integration variable and the external momentum $x$ component in Eq.\eqref{eq:aDa} through $q=k_2 v/\gamma$ and $k=k_1 v/\gamma$. This is equivalent to choosing the unit of length equal to the mean free path $l_{\rm ee}=v/\gamma$. Recalling that $\sinh\beta=\frac{\gamma}{|\vec k|v}$ we write
\be\label{eq:aDa_2}
\hat\alpha D\hat\alpha=\frac{\alpha^2}{2\pi v}\lp\begin{array}{cc} I_1(k) &-I_2(k)\\  -I_2(k) & I_1(k) \end{array}\rp
,
\ee
where 
\be
I_1(k)=\int\limits_{-\infty}^\infty \frac{dq e^{\beta}}{\sqrt{q^2+k^2}}
,\quad
I_2(k)=\int\limits_{-\infty}^\infty \frac{dq e^{\beta} z_k^2}{\sqrt{q^2+k^2}}
.
\ee
Straightforward integration gives 
\be
\begin{split}
& I_1(k)=\frac{2}{|k|}\lp \pi-\tan^{-1}|k|\rp+2\ln\frac{2 q_{\rm uv}}{a}
\\
& I_2(k)=2+2k\tan^{-1}\frac1{k}-2\ln\frac{2 q_{\rm uv}}{a}
\end{split}
\ee
where $a=\sqrt{k^2+1}$ and $q_{\rm uv}$ is the UV cutoff set by the microscopic width  
of the 
line scatterer at $y=0$. 

Plugging these results in Eq.\eqref{eq:df} we obtain
\be\label{eq:df_1}
\left. | f\ra=\lp G-G\frac{\alpha P'}{1+\frac{\alpha}{2\pi v} M(k)}D
\rp \left. J_0|0\ra
\ee
where we defined a matrix
\be
M(k)=\lp\begin{array}{cc} I_1(k) &-I_2(k)\\  -I_2(k) & I_1(k) \end{array}\rp
.
\ee
The first term  in Eq.\eqref{eq:df_1} describes point source in free space, the second term describes the result of multiple encounters with the line $y=0$. 
To understand its structure we note that the quantity $P' D\left. |0\ra$ is given by the first and last terms in 
\be\label{eq:f0}
D\left.|0\ra= \frac{e^{i\theta}\left.|1\ra}{iv(k_1+ik_2)}+\frac{e^{-\beta}\left.|0\ra}{vk}+\frac{e^{-i\theta}\left.|-1\ra}{iv(k_1-ik_2)}
\ee
integrated over $k_2$ (here $e^{-\beta}=\sqrt{1+\frac{\gamma^2}{k^2v^2}}-\frac{\gamma}{kv}$). Using the Cauchy integral 
$\int\limits_{-\infty}^\infty \frac{dq}{2\pi(k\pm iq)}=\frac12\sgn k$
we find that the $m=\pm 1$ components are equal in magnitude and in sign. Namely, $P'G\left. |0\ra\sim \lp\begin{array}{c} 1\\1\end{array}\rp$ which is an eigenvector of the matrix $M(k)$ in Eq.\eqref{eq:df_1} with the eigenvalue $I(k)=I_1(k)-I_2(k)$. We can therefore replace Eq.\eqref{eq:df_1} with
\be\label{eq:df_2}
\left. | f\ra=G\lp\begin{array}{c} w_k\\1\\w_k\end{array}\rp 
J_0
,\quad
w_k=\frac{i\alpha\,\sgn k}{2v(1+\frac{\alpha}{2\pi v}I(k) )}
\ee
where the $m=\pm1$ components $w_k$ generate the line contribution whereas the $m=0$ component (equal to $1$) generates the point-source contribution. 
This expression is used in the main text to analyze the response measured by a potential probe at the boundary. 

\end{document}